\documentclass[aps,pra,amsmath,amsfonts,floatfix,superscriptaddress,notitlepage,nofootinbib]{revtex4-1}

\usepackage{graphicx,graphics}
\usepackage{epstopdf}
\usepackage{amsmath}
\usepackage{amssymb}
\usepackage{bbold}
\usepackage[colorlinks=true]{hyperref}

   \let\L=\Lambda

\def\beq{\begin{equation}}
\def\eeq{\end{equation}}

\def\DD{\mathcal{D}}

\renewcommand\b[1]{\boldsymbol{#1}}

\renewcommand\bar[1]{\overline{#1}}

\def\to{\rightarrow}

\def\tr{\text{Tr}}
\def\tr{\text{tr}}
\def\str{\text{str}}
\def\Str{\text{Str}}

\def\Q{\mathcal Q}

\begin{document}

\title{Exact generating function of a zero-dimensional supersymmetric non-linear sigma model}
\author{Adam Ran\c con}
\address{Universit\'e de Lille, CNRS, UMR 8523 -- PhLAM -- Laboratoire de Physique des Lasers Atomes et Mol\'ecules, F-59000 Lille, France}
 \author{Ivan Balog}
\address{Institute  of  Physics,  Bijeni\v cka  cesta  46,  HR-10001  Zagreb,  Croatia}

\begin{abstract}
We compute exactly the generating function of a supersymmetric non-linear sigma model describing random matrices belonging to the unitary class.  Although an arbitrary source explicitly breaks the supersymmetry, a careful analysis of the invariance of the generating function allows us to show that it depends on only three invariant functions of the source. 
This generating function allows us to recover various results found in the literature. It also questions the possibility of a functional renormalization group study of the three-dimensional Anderson transition.
\end{abstract}

\maketitle

\section{Introduction}

Supersymmetry is a powerful tool for the study of random systems, ranging from random matrices theory (RMT), quantum chaotic  and disordered systems \cite{EfetovBook,HaakeBook,Mirlin2000}, with connection to string theory and the SYK model \cite{Altland2020,Monteiro2020,Sedrakyan2020}. In particular, it has proven to be most useful in the study of Anderson localization in the weak disorder limit. The corresponding supersymmetric Non-Linear Sigma model  (SUSY NLSM) pioneered by Efetov is heuristically given by the effective field theory
\begin{equation}
Z= \int \DD Q\,e^{-S[Q]},
\end{equation}
where the supermatrix field $Q$ lives on some target space (see below), $S[Q]=\int d^dx\, \Str(\nabla Q )^2+S_{\rm br}[Q]$ and $S_{\rm br}[Q]$ breaks partially the supersymmetry \cite{EfetovBook}.
 In the zero-dimensional (or zero-mode limit), where the spatial fluctuations of $Q$ are neglected, this theory describes random matrices, quantum chaotic systems in the ergodic regime, and disordered metallic grains \cite{EfetovBook,HaakeBook}. In one-dimension, this method allows for a very fine description of Anderson localization and its dynamics \cite{Efetov1983a,Micklitz2014,Khalaf2017,Khalaf2017a}. 

On the other hand, concerning the Anderson transition,  which is known to exist in dimension greater than two (in the unitary class on which we focus here), the picture is much less appealing. The SUSY NLSM does find a transition on the Bethe lattice \cite{Efetov1987,Zirnbauer1986,Zirnbauer1986a,Mirlin1991} and in $2+\epsilon$ dimensions \cite{EfetovBook}, but those are very specific results that are hard to extrapolate to the more sensible three-dimensional case. Indeed, close to the lower critical dimension, the SUSY NLSM reproduces the perturbative renormalization group results, which are however known to be badly behaved and cannot reasonably be extrapolated to large dimensions. The standard remedy is usually to expand close to the upper critical dimension, which tends to give better behaved asymptotic series. The Anderson transition is in this respect peculiar since this dimension is suspected to be infinite, and to be described by the transition on the Bethe lattice. However, the critical properties in that case are rather peculiar and different from the transition expected on a hypercubic lattice. 
 Therefore, a field-theoretic description of the Anderson transition has to be non-perturbative in essence.

A promising avenue is the functional renormalization group (FRG), a modern implementation of Wilson ideas \cite{Berges2002,Dupuis2020}. An essential feature, which allows for non-perturbative approximations of the renormalization group flow equations, is that it is functional in the fields, $Z\to Z[J]$ with $J$ a (space dependent) source term that completely breaks the supersymmetry.  However, to our knowledge, the functional form of the generating function of the SUSY NLSM, needed for an implementation of the FRG, has not been studied for generic sources completely breaking the supersymmetry, even in the simplest case of the zero-dimensional limit. 
Due to the supersymmetry and the non-linear constraints of this field theory, the field dependence can be expected to be rather complicated. This is what we explore here in the zero-dimensional limit, that is, we compute the generalized generating function
\begin{equation}
Z[J]= \int \DD Q\,e^{\Str(J Q)},
\end{equation}
where the supermatrix source $J$ is arbitrary. From a RMT perspective, this generating function corresponds to the generalized spectral determinant
\begin{equation}
Z[J]=\langle{\rm Sdet} (J-H)\rangle_{\rm GUE},
\end{equation}
where the average is done over the hermitian matricess $H$ belonging to the Gaussian Unitary Ensemble (GUE). For diagonal source, one recovers the standard spectral determinant which allows for the calculation of  spectral properties of the GUE universality class.

This manuscript is organized as follows. In Section \ref{sec_defZ}, we define the generating function of the SUSY NLSM for an arbitrary source. We also analyze the constraints imposed on the source to ensure convergence of the superintegrals, as well as its symmetries and invariance properties. In Section \ref{sec_compZ}, we give the explicit form of the generating function and discuss its functional dependence, and in Section \ref{sec_oldres}, we show how our calculation allows us to recover various results of the literature. We discuss our results and future works in Section \ref{sec_concl}.

\section{Generating function  : definition, symmetries and invariance\label{sec_defZ}}

\subsection{Definition of the generating function}

We aim at computing the generating function of the zero-dimensional SUSY NLSM corresponding to the GUE, which is defined as
\begin{equation}
Z[J]= \int \DD Q\,e^{\Str(J Q)},
\end{equation}
with $J$ a supermatrix source. The construction of this model, including the definition of the measure $\DD Q$, can be found for example in \cite{HaakeBook}. The supermatrix $Q$ is written as $Q=T\L T^{-1}$ where $T$ belongs to the coset space $U(1,1|2)/U(1|1)\times U(1|1)$, and  $\Lambda={\rm diag}(1,1,-1,-1)$. Here, $U(1,1|2)$ is the group of pseudo-unitary supermatrices that leaves the matrix $K={\rm diag}(1,1,1,-1)$ invariant, $UK U^\dagger=K$ for all $U\in U(1,1|2)$, while $U(1|1)\times U(1|1)$ is its subgroup of pseudo-unitary matrices that commute with $\L$. The non-linear nature of the model is reflected in the fact that $Q^2=\mathbb{1}$. We note, as it will be important in the following, that $\Str(Q)=0$.

The source $J$ allows for the computation of any correlation function of the supermatrix $Q$ (in presence of the source). The main difficulty is that the source completely breaks the supersymmetry $Q\to UQU^{-1}$,  $U\in U(1,1|2)$, which usually allows for very powerful supersymmetric integration theorems \cite{WegnerBook}. However, the existence of this symmetry (in the absence of source),  as well as others that we discuss below, implies some invariances of the generating function. It is the exploitation of these invariances that will allow us to compute $Z[J]$ for arbitrary sources.

Before continuing, we give our conventions. We mostly follow those of Efetov in \cite{EfetovBook}, and we spell out the ones that will be important in the following. The $4$-by-$4$ supermatrices above operate in the advanced-retarded ($AR$) space and the Fermi-Bose ($FB$) superspace. The supertrace is defined as $\Str\, Q = \str \,Q^{AA}+ \str\, Q^{RR}$ with $ \str\, Q^{\alpha\alpha'}=Q^{\alpha\alpha'}_{FF}-Q^{\alpha\alpha'}_{BB}$ (${\rm Sdet(\ldots)}=\exp(\Str\log(\ldots))$. Here $Q^{\alpha\alpha'}_{FF}$ and $Q^{\alpha\alpha'}_{BB}$ are complex numbers while $Q^{\alpha\alpha'}_{FB}$ and $Q^{\alpha\alpha'}_{BF}$ are Grassmann variables. In the following, $\sigma^\nu$ and $\tau^\nu$, $\nu=0,\ldots,3$ correspond to the Pauli matrices acting respectively in the $AR$ and $FB$ sectors ($\nu=0$ corresponding to the identity matrix). 

In the following, we will use the rational parametrization of $Q$ \cite{HaakeBook},
\begin{equation}
Q=\begin{pmatrix}
\tau_0 && W\\
\overline W && \tau_0
\end{pmatrix}\begin{pmatrix}
\tau_0 && 0\\
0 && -\tau_0
\end{pmatrix}\begin{pmatrix}
\tau_0 && W\\
\overline W && \tau_0
\end{pmatrix}^{-1},
\end{equation}
 and
\begin{equation}
\begin{split}
W&=\begin{pmatrix}
x && \nu\\
\mu  && y
\end{pmatrix},\\
\overline W &=\begin{pmatrix}
-x^* && \bar\mu\\
 \bar\nu && y^*
\end{pmatrix}
\end{split},
\end{equation}
with $x$ and $y$ two complex variables, and $\mu$, $\nu$, $\bar\mu$, $\bar\nu$ four Grassmann variables. One shows that
\begin{equation}
 Z[J]=-\int \DD W\DD \tilde{ W} e^{\Str(LQ)},
\end{equation} 
with
\begin{equation}
\DD W\DD \overline W=\frac{d^2 x}{\pi}\frac{d^2 y}{\pi}d\bar\mu\, d\mu\, d\bar\nu\,  d\nu,
\label{eq_measureW}
\end{equation}
and the range of the complex variables is $|x|\in  [0,\infty[$, $|y|\in [0,1]$.

\subsection{Convergence}
The generating function Z[J] is defined only if the integral over $Q$ is convergent. Because of the non-compact nature of the bosonic sector (associated to $y$), this is not possible for arbitrary values of the $16$ matrix elements of the source. It is convenient to parametrize the source as $J=U_0 L U_0^{-1}$, with $L={\rm diag}(l^A_{F},l^A_{B},l^A_{F},l^R_{B})$ and $U_0\in U(1,1|2)$, using that $U_0$ depends on $8$ fermionic and $4$ bosonic variables, and that $L$ gives $4$ additional bosonic degrees of freedom. Thanks to the invariance of the measure $\DD Q$ under $Q\to U_0 Qu_0^{-1}$, the generating function can be rewritten in terms of the eigenvalues of the source $l^s_\alpha$,
\begin{equation}
Z[J]=\int \DD Q\, e^{\Str(L Q)}.
\label{eq_ZL}
\end{equation}
 Using the rational parametrization given above, one finds that the purely bosonic part of $\Str(L Q)$ is
\begin{equation}
\Str(L Q) = \frac{1-|x|^2}{1+|x|^2}(l_{F}^A-l_{F}^R)- \frac{1+|y|^2}{1-|y|^2}(l_{B}^A-l_{B}^R),
\end{equation}
which implies the constraint $l^A_{B}>l^R_{B}$ for convergence.

In the $FB$ notation, $\Str(J Q)=\tr(J_{FF}Q_{FF})+\tr(J_{FB}Q_{BF})-\tr(J_{BF}Q_{FB})-\tr(J_{BB}Q_{BB})$. Due to the compact nature of $Q_{FF}$, and the Grassmann nature of $Q_{FB}$ and $Q_{BF}$, the corresponding terms will always give convergent contributions. As usual, 
the dangerous part comes only from the $BB$ sector. Forgetting for simplicity about the Grassmann contribution to $Q_{BB}$ (which does not change the convergence properties), we can write it as
\begin{equation}
Q_{BB}=\begin{pmatrix}
\frac{1+|y|^2}{1-|y|^2}  && \frac{y}{1-|y|^2}\\
-\frac{y^*}{1-|y|^2}&& -\frac{1+|y|^2}{1-|y|^2}
\end{pmatrix},
\end{equation}
which can be written as $ u.{\rm diag}(1,-1).u^{-1}$, with $u\in SU(1,1)/U(1)$. It is therefore convenient to parametrize $J_{BB}$ such that it can be diagonalized using $SU(1,1)$ matrices, for instance
\begin{equation}
J_{BB}=\begin{pmatrix}
b  && if\\
if^*&& -d
\end{pmatrix},
\label{eq_JBB}
\end{equation}
with the constraint that the real part of $(b+d)-2\sqrt{ff^*}$ is positive. This insures that for its eigenvalues that $l^A_{B}>l^R_{B}$ and thus convergence. In practice, $f$ and $f^*$ are treated as independent variables, with the only constraint that the integral converges.

On the other hand, $Q_{FF}$ is diagonalizable by $SU(2)$ matrices, and it is therefore convenient to write $J_{FF}$ as an Hermitian matrix
\begin{equation}
J_{FF}=\begin{pmatrix}
a  && e\\
e^*&& c
\end{pmatrix}.
\end{equation}

\subsection{Symmetries and invariance}

\subsubsection{$U(1,1|2)$ supersymmetry}
The symmetries of the measure $\DD Q$ under a given group transformation implies an invariance of the generating function under any transformation of the source by the same group. For instance, the measure is by construction invariant under  $Q\to U^{-1} Q U$ with $U\in U(1,1|2)$, which implies that 
\begin{equation}
Z[J]=Z[U J U^{-1}], \quad  \forall\; U\in U(1,1|2).
\end{equation}
This invariance under a continuous transformation is highly constraining, since it implies that $Z[J]$ only depends on the source through $\Str\left(J^n\right)$, $n\geq 1$. Furthermore, since the $\Str\left(J^n\right)$ depends only on the four eigenvalues of $J$, only four of these supertraces are independent (this is an equivalent of the Cayley-Hamilton theorem for ordinary square matrices). Therefore, the generating function is at most a function of the four  $U(1,1|2)$ invariant variables $j_n=\Str\left(J^n\right)$, $n=1,\ldots,4$.

For strictly diagonal sources, another discrete symmetry, the so-called Weyl symmetry, is known to exist, see e.g.  \cite{HaakeBook} or \cite{Altland2020}. In our notations, it amounts to the exchange $l^A_F \leftrightarrow l^R_F$. However, this symmetry is included in $U(1,1|2)$, as it corresponds to a transformation $Q\to U_W^{-1} Q U_W$ with $U_W=u_{FF}\otimes\mathbb{1}_{BB}$ and $u_{FF}=e^{i\frac{\pi}{2}\sigma^2_{FF}}$, with $\sigma^2_{FF}$ the second Pauli matrix acting in the $FF$ subspace only. This discrete Weyl transformation can therefore been made continuous by choosing an arbitrary $u_{FF}\in SU(2)$. If $J$ is block-diagonal in the $FB$ representation (that is, if $J_{FB}=J_{BF}=0$), this implies that the fermionic block $J_{FF}$ is transformed independently $J_{FF}\to u_{FF}J_{FF}u_{FF}^{-1}$, with the corresponding invariance of the generating function. With this insight, we can therefore define a generalized continuous Weyl transformation for a $FB$-block-diagonal source that also transforms the bosonic sector, with $U_W\in SU(2)\otimes SU(1,1)$. Note that due to the non-compact nature of the bosonic sector, this does \emph{not} lead to an exchange $l_{BB}^{AA} \leftrightarrow l_{BB}^{RR}$ for diagonal sources.

\subsubsection{$AR$ exchange symmetry}

There exists however another discrete invariance, to our knowledge not discussed in the literature, that in essence exchanges the advanced and retarded component of the source. With the rational parametrization given above, this transformation amounts to the change of variables $W\leftrightarrow\overline W$
of unit Jacobian, and under which $Q\to -E Q E$, with $E_{\alpha\alpha'}^{ss'}=\sigma^1_{ss'} \delta_{\alpha\alpha'}$. That is, this change of variable exchanges the advanced and retarded components of $Q$, up to a sign,
\begin{equation}
Q=\begin{pmatrix}
Q^{AA}  && Q^{AR}\\
Q^{RA}  && Q^{RR}
\end{pmatrix}\to-\begin{pmatrix} 
Q^{RR}  && Q^{RA}\\
Q^{AR}  && Q^{AA}
\end{pmatrix}.
\end{equation}
The generating function is thus invariant under the corresponding transformation of the source $Z[J]=Z[-EJE]$. Note that this exchange of advance and retarded components of the source does not compromise the convergence of the integral, thanks to the minus sign. For diagonal sources, this transformation does  lead to an exchange in the bosonic sector $l_{BB}^{AA} \leftrightarrow- l_{BB}^{RR}$, contrary to the generalized Weyl invariance.

Under this transformation, one readily sees that the $U(1,1|2)$ invariants transform as $j_n\to (-1)^n j_n$.

\subsubsection{Shift invariance}

There is one additional invariance of the generating function, not due to a symmetry of the measure, but to the constraint $\Str(Q)=0$. This implies directly that $Z[J]=Z[J+t\mathbb{1}]$, for all $t$. This invariance is much more promising that what might appear at first sight, as it constraints the generating function to depend on only three combinations of the $U(1,1|2)$ invariants $j_n$.

Collecting the invariants in a vector, $\b{j}=(j_1,j_2,j_3,j_4)$, an explicit calculation shows that under a shift $J\to J+t\mathbb{1}$, they transform as
\begin{equation}
\b j(t)= M_t\b j,
\end{equation}
with
\begin{equation}
M_t=\begin{pmatrix}
1 & 0 & 0 &0\\
2t & 1 & 0 & 0\\
3t^2 & 3t & 1 &0\\
4t^3 & 6t^2 & 4t & 1
\end{pmatrix}.
\end{equation}
It is convenient to rewrite this matrix as $M_t=e^{S t}$, with 
\begin{equation}
S=\begin{pmatrix}
0 & 0 & 0 &0\\
2 & 0 & 0 & 0\\
0 & 3 & 0 &0\\
0 & 0 & 4 & 0
\end{pmatrix},
\end{equation}
the generator of the transformation, making explicit that this transformation is a one-parameter Lie group. The theory of such groups allows the construction of invariant functions \cite{CohenBook}. Infinitesimal transformations are generated by the differential operator $D=\b j S\partial_{\b{j}}=2j_1 \partial_{j_2}+3j_2 \partial_{j_3}+4j_3 \partial_{j_4}$, i.e. $\partial_t \b j(t)=D\b j(t)$, which is solved by
\begin{equation}
\begin{split}
i_1(\b j(t))&\equiv j_1(t)=c_1,\\
i_2(\b j(t))&\equiv j_1(t)j_3(t)-\frac34 j_2(t)^2=c_2,\\
i_3(\b j(t))&\equiv j_1(t)^2 j_4(t)-2 j_1(t) j_2(t) j_3(t)+j_2(t)^3=c_3,\\
v(\b j(t))&\equiv\frac{j_2(t)}{2 j_1(t)}=c_4+t,
\end{split}
\end{equation}
with $c_1,\ldots,c_4$ four constants. Any function $f(\b j)$ can be written as a function $f(i_1,i_2,i_3,v)$ of the three invariants $i_n$, $n=1,2,3$, and $v$ which varies linearly with $t$. Therefore, a function of $\b j$  invariant under the shift transformation is a function of  three invariants $i_1$, $i_2$ and $i_3$ only.
Note that while $i_2$ and $i_3$ are invariant under the advanced-retarded exchange symmetry, $i_1$ is not. This is easily corrected by redefining it as $i_1\equiv j_1^2$.

To summarize, thanks to the $U(1,1|2)$, advanced-retarded exchange, and shift invariances, we have shown that in all generality, the generating function $Z[J]$, in appearance a function of $16$ variables, is really just a function of three invariants, $Z[i_1,i_2,i_3]$, with
\begin{equation}
\begin{split}
i_1&=\Str(J)^2,\\
i_2&= \Str(J)\Str\big(J^3\big)-\frac34\, \Str\big(J^2\big)^2,\\
i_3&=\Str(J)^2 \Str\big(J^4\big)-2\, \Str(J)\Str\big(J^2\big)\Str\big(J^3\big)+\Str\big(J^2\big)^3.
\end{split}
\label{eq_ItoJ}
\end{equation}

\section{Calculation of the generating function \label{sec_compZ}}

\subsection{Explicit form of the generating function \label{sec_calcZ}}
We are now in a position to compute the generating function for arbitrary source $J$. First, using the $U(1,1|2)$ invariance, we start by diagonalizing the source $J\to U_0^{-1}JU_0= L={\rm diag}(l^A_{F},l^A_{B},l^R_{F},l^R_{B})$, and compute the generating function in terms of the four eigenvalues of the source.\footnote{In \cite{HaakeBook}, the integration over $Q$ is performed by using a change of variable corresponding to diagonalizing $W\overline W$ and $\overline W W$, similar in spirit to Efetov's parametrization \cite{EfetovBook}. This transformation introduces a singular Berezinian that needs to be dealt with. In \cite{HaakeBook}, the Efetov-Wegner term is not computed, and the generating function is obtained by imposing the discrete Weyl symmetry.}
Performing the integral over the Grassmann variables explicitly,  one is left with the integrals over  $|x|$ and $|y|$ after a trivial integration over their arguments. The remaining integrals can be performed using the changes of variable $|x|=\sqrt{\frac{1-r_F}{1+r_F}}$ and $|y|=\sqrt{\frac{r_B-1}{r_B+1}}$, and one obtains
\begin{equation}
\begin{split}
Z[L]&=Z_0[L]+Z_{1}[L],\\
Z_0[L]&=\frac{e^{l_F^A-l_F^R-l_B^A+l_B^R} 
   \left(l_F^A-l_B^R\right) \left(l_B^A-l_F^R\right)}{\left(l_B^A-l_B^R\right)
   \left(l_F^A-l_F^R\right)},\\
 Z_1[L]&=\frac{e^{-l_F^A+l_F^R-l_B^A+l_B^R}
   \left(l_F^A-l_B^A\right)
   \left(l_F^R-l_B^R\right)}{\left(l_B^A-l_B^R\right)
   \left(l_F^A-l_F^R\right)}.
\end{split}
\label{eq_Zdiag}
\end{equation}
It is known that for the unitary class, the semiclassical limit -- the evaluation of the integral as a saddle point analysis and gaussian fluctuations -- is in fact exact, see e.g. \cite{Zirnbauer1999,Altland2020}. What is needed is to take into account two saddles, the standard one $Q_0=\Lambda$, as well as the Andreev-Altshuler saddle \cite{Andreev1995}, which corresponds to the Weyl transformation of the standard saddle $Q_1=U_W^{-1} Q_0 U_W={\rm diag}(-1,1,1,-1)$. In particular, it is straightforward to show that expanding $Q$ to quadratic order in $W$ and $\overline W$ in Eq.~\eqref{eq_ZL} and performing the integral over $W$ and $\overline W$, one obtains $Z_0[L]$. The other contribution $Z_1[L]$ can be obtained by performing the semi-classical analysis around $Q_1$, or directly using the Weyl symmetry, as $Z_1[L]=Z_0[U_W L U^{-1}_W]$.

 To rewrite the generating function in terms of the invariants $i_1$, $i_2$ and $i_3$, it is necessary to write the combinations of $l^A_{F}$, $l^A_{B}$, $l^R_{F}$ and $l^R_{B}$ that appear in Eq.~\eqref{eq_Zdiag}  in terms of the invariants. Since we have four eigenvalues, but only three invariants, this relationship cannot be inverted directly.\footnote{In principle, it is possible to rewrite the eigenvalues in terms of the four $U(1,1|2)$ invariants $j_n$. In addition to losing the explicit shift-invariance, this approach implies to solve four equations up to quartic order, which is hard to handle.}
A much more convenient approach is to rewrite the diagonalized source  as $L=I_F e_F+I_B e_B+j_1 e_S+x \mathbb{1}$, with the basis 
\begin{equation}
\begin{split}
 e_F&=\frac12{\rm diag}(1,0,-1,0),\\
 e_B&=\frac12{\rm diag}(0,1,0,-1),\\
 e_S&=\frac14{\rm diag}(1,-1,1,-1),
 \end{split}
 \label{eq_basis}
\end{equation}
The shift invariance implies that  the generating function depends on $I_F$, $I_B$ and $j_1$   only (here $j_1$ does correspond to $\Str(J)$ as can be checked directly). 
With this parametrization (using $j_1^2=i_1$), we obtain
\begin{equation}
\begin{split}
Z_0[J]&=e^{I_F-I_B}\frac{ (I_B+I_F)^2-i_1}{4I_F I_B},\\
 Z_1[L]&=-e^{-I_F-I_B}\frac{ (I_B-I_F)^2-i_1}{4I_F I_B}.
\end{split}
\label{eq_Z0_Z1}
\end{equation}
and the generating function becomes
\begin{equation}
Z[J]=e^{-I_B}\left(\cosh(I_F)+\frac{I_F^2+I_B^2-i_1}{2 I_F I_B}\sinh(I_F)\right).
\label{eq_Zfinal}
\end{equation}
Convergence of the superintegral translates here into the constraint $I_B>0$, and we note that the generating function is independent of the sign of  $I_F$.  Writing the invariants $i_2$ and $i_3$ in terms of $I_F$, $I_B$ and $i_1$, and solving for $I_F$ and $I_B$, one obtains
\begin{equation}
\begin{split}
I_F &=\sqrt{\frac{9 i_3^2}{i_1\left(4 i_2-i_1^2\right)^2}+\frac{4
   i_2}{3i_1}-\frac{i_1}{12}-\frac{3 i_3}{4 i_2-i_1^2}},\\
I_B &=\sqrt{\frac{9 i_3^2}{i_1\left(4 i_2-i_1^2\right)^2}+\frac{4
   i_2}{3i_1}-\frac{i_1}{12}+\frac{3 i_3}{4 i_2-i_1^2}}.
\end{split}
\label{eq_IBIF}
\end{equation}
Writing $I_B$ and $I_F$ in terms of $j_n$ does not simplify these expressions. While their full source dependence is rather complicated, it becomes much nicer for sources that depends purely on bosonic variables, that is, for $J_{FB}=J_{BF}=0$. In this case, we have
\begin{equation}
\begin{split}
I_B&=\sqrt{2\tr(J_{BB}^2)-\tr(J_{BB})^2},\\
I_F&=\sqrt{2\tr(J_{FF}^2)-\tr(J_{FF})^2},\\
i_1&=\big(\tr(J_{FF})-\tr(J_{BB})\big)^2.
\end{split}
\label{eq_Ibos}
\end{equation}

Equation \eqref{eq_Zfinal} is our main result, from which one can derive various results found in the literature. Note that our expression, even if it does not seem so, is especially compact. Trying to write it in terms of the matrix elements of the source gives rise to extremely large expressions, that are not easily handled, even with the use of computer algebra system like Mathematica. As we will discuss below, it seems rather difficult to see how a brute force calculation of $Z$ for an arbitrary source could be done and written in terms of invariants.

\subsection{Expansion in terms of Grassmann variables}

One stringent check of Eq.~\eqref{eq_Zfinal} consists in expanding it in terms of the Grassmann matrix elements of the source. First of all, the generating function has even grading, which implies that the expansion generates products of even numbers of Grassmann matrix elements of the source. Furthermore, noting that the matrix $Q$ only depends on four Grassmann variables, one realizes that no products of six or height  Grassmann matrix elements can be generated. Given Eqs.~\eqref{eq_Zfinal} (in particular in view of Eqs.~\eqref{eq_ItoJ} and \eqref{eq_IBIF}), it does not seem obvious at all that it would be so. By carefully expanding the generating function is powers of the Grassmann matrix elements of the source (which generates thousands upon thousands terms), we have checked explicitly that all terms involving six or more components of the source exactly compensate each other, such that only terms with product of zero, two or four Grassmann matrix elements survive.

We also compared our expansion to a brute force calculation of the generating function with an arbitrary source. This expansion was done using Efetov's parametrization \cite{EfetovBook}, with the integrals over the Grassmann variables performed explicitly. The remaining integrals over the bosonic variables could be performed explicitly in some cases (for all terms with four Grassmann matrix elements and some terms with two Grassmann matrix elements), which reproduce exactly the result obtained by expanding Eq.~\eqref{eq_Zfinal}. In the remaining cases, we checked numerically that the numerical integrations reproduce our result for various numerical values of the source.

Note that even if we were able to compute all integrals explicitly, it seems nearly impossible to guess the resummation of the Grassmann matrix elements of the source in a way that would reproduce Eq.~\eqref{eq_Zfinal}.

\subsection{Functional dependence of the generating function}

The dependence of the generating function, Eq.~\eqref{eq_Zfinal}, on the source is rather complicated. We start with a few observations. First of all, it is singular in the limit $I_B\to 0$ due to the non-compact nature of the bosonic sector (recall that $I_B>0$ for the superintegral over $Q$ to be defined). Second, for a source which eigenvalues are of the form ${\rm diag}(a,a,b,b)$ (corresponding to $I_F=I_B$ and $i_1=0$), one finds $Z=1$ for all $a>b$. This is due to the exact  $U(1|1)\times U(1|1)$  supersymmetry which is left unbroken by the source \cite{Zirnbauer1986a}. Finally, we note that the generating function is allowed to change sign. This is of course due to the mixture of bosonic and fermionic variables in the superintegral, which implies that the generating function cannot be interpreted as the partition function of a standard Boltzmann distribution.

While none of these facts are in themselves problematic for the generating function, they raise some questions concerning the functionals that are at the center of quantum field theory and renormalization group methods. A simple example is of course the generating function of the connected correlation functions, $W[J]\equiv \ln Z[J]$, which is singular when $Z[J]$ vanishes. Another important quantity, especially in the FRG formalism, is the effective action, defined as the Legendre transform of $W[J]$ with respect to $J$. In order to define it, one must first compute $\Q[J]=\langle Q\rangle_J$, the average of $Q$ as a function of the source, and then invert the relationship between $\Q$ and $J$ to obtain $J[\Q]$. Finally, the effective action is then defined by $\Gamma[\Q] = -W[J[\Q]]+\Str(\Q J[\Q])$.

Thanks to the $U(1,1|2)$ invariance of the generating function, the effective action is expected to also be invariant under $\Q\to U^{-1}\Q U$ with $U\in U(1,1|2)$, while the linear constraint on the field $\Str (Q)=0$ implies $\Str (\Q[J])=0$ for arbitrary sources. All this implies that one should parametrize $\Gamma[\Q] $ in terms of the three invariants $q_n=\Str(\Q^n)$, $n=2,3,4$.\footnote{The exchange transformation implies that $\Gamma$ should depend on $q_3^2$. We also stress that the non-linear constraint $Q^2=\mathbb{1}$ does \emph{not} imply $\Q[J]^2=\mathbb{1}$, see for instance \cite{Rancon2019}.} The calculation of the invariants $q_n$ is not particularly enlightening, and is left to appendix \ref{app_Q}.

From the explicit calculation of $\Q[J]$, we observe that for  sources such that $j_1=\pm (I_F-I_B)$, one has all $q_n=0$. This corresponds to having sources with equal eigenvalues in the advanced or retarded sectors (or in both sectors). This is in agreement with (but slightly more general than) the expected $\Q[J=\alpha \Lambda]=\Lambda$, for physical values of the source $J \propto \Lambda$. In particular, this implies that we cannot invert the relationship between $\Q$ and $J$ in general. In agreement with this, we find that these kinds of sources, the would-be effective action $\hat \Gamma[J]=-W[J]+\Str(\Q[J] J)$ always vanishes. Once again, these results are related to the $U(1|1)\times U(1|1)$  supersymmetry, which is only partly broken for these sources. We therefore conclude that the supersymmetry inherent to this model, which makes it particularly powerful in other contexts, does not allow for the zero-dimensional effective action to exist.

While our calculation is explicit only for the zero-dimensional effective action, the reasoning can be generalized to any dimensions, as long as the source is constant in position and does not completely break the global $U(1|1)\times U(1|1)$  supersymmetry. Indeed, by the same argument, we expect this kind of source to generate an homogenuous $\Q[J]$ such that $\Str(\Q^n)=0$.  Then the effective potential (i.e. the effective action in constant field) does not exist. While in this case, we cannot exclude that the effective action does exist (as long as the source is not constant), it nevertheless questions the use of the FRG schemes based on the effective action (a la Wetterich \cite{Berges2002}) to address the Anderson transition.

\section{Recovering various results from the literature \label{sec_oldres}}

The calculation of the generating functional allows us to recover various results found in the literature straight-forwardly. Any correlation function of the matrix $Q$ can be obtained by taking derivatives of the generating function with respect to the source. They are usually obtained by computing directly the correlation function for a physical value of the source (proportional to $\Lambda$, see below), and therefore, each necessitates a separate, sometimes rather tedious, calculation. With our approach, the most difficult part has been delegated to the calculation of the generating function for diagonal source, and the remaining calculation of correlation functions is then much easier.

\subsection{Level correlation function}
One of the simplest and most iconic correlation function of random matrix theory is the level correlation function of the Gaussian Unitary Ensemble $R(s)$ \cite{MehtaBook}, where $s$ is the energy in units of the mean level spacing (it is also related to the mesoscopic echo of a quantum dot \cite{Prigodin1994}). It is given in the NLSM by 
\begin{equation}
R(s) = -\frac12 {\rm Re}\langle Q_{FF}^{AA} Q_{FF}^{RR}\rangle_{c,J_s},
\end{equation}
for a source $J_s=-i \frac s2 \Lambda$, and the subscript $c$ means that we only keep the connected part (see for instance \cite{Altland2020}). This correlation function can be obtained from the generating function for purely diagonal source (as discussed in \cite{HaakeBook}),
\begin{equation}
R(s) = -\frac12 {\rm Re} \frac{\delta^2 \log Z}{\delta J^{AA}_{FF}\delta J^{RR}_{FF}}\bigg|_{J_s},
\end{equation}
from which ones readily obtained the well-known result
\begin{equation}
R(s) = -\frac{1-\cos^2 s }{s^2}.
\end{equation}
As is well known, the non-oscillatory contribution comes from perturbative expansion around the standard saddle-point $Q_0=\Lambda$, while the non-perturbative oscillatory term can be obtained from an expansion around the Andreev-Altshuler saddle $Q_1$ \cite{Andreev1995a} (see discussion in Sec.~\ref{sec_calcZ}).

\subsection{Probability distribution of the density of state}
From the generating function, we can compute much more demanding observables.
One is  the probability distribution of density of state of a metallic grain.
Following Efetov \cite{Efetov1993}, the probability distribution of the density of state is given by
\begin{equation}
P_\rho(s)=\int_0^{2\pi}\frac{dt}{2\pi}\left\langle \delta\left(s-\frac12\left(Q^{AA}_{BB}-Q^{RR}_{BB}+e^{it}Q^{AR}_{BB}-e^{-it}Q^{RA}_{BB}\right)\right)\right\rangle_{J_0},
\end{equation}
where the physical value of the source is $J_0=\frac\gamma2\Lambda$. Using that $Z[J_0]=1$, one shows easily that the Laplace transform of $P_\rho(s)$, $\tilde P_\rho(k)=\int_0^\infty e^{-k s}P_\rho(s)$ reads
\begin{equation}
\tilde P_\rho(k)=\int_0^{2\pi}\frac{dt}{2\pi}Z[J_0+J_1],
\end{equation}
with 
\begin{equation}
(J_1)_{BB}=\frac k2\begin{pmatrix}
1 & -e^{-it} \\
e^{it} & -1
\end{pmatrix},
\end{equation}
and the other submatrices of $J_1$ being zero. Using Eqs.~\eqref{eq_Zfinal} and \eqref{eq_Ibos}, we have 
\begin{equation}
\begin{split}
\tilde P_\rho(k)&=\left(\cosh (\gamma )+\sinh (\gamma ) \frac{\gamma +k}{\sqrt{\gamma  (\gamma +2 k)}}\right)e^{-\sqrt{\gamma  (\gamma +2 k)}},\\
&=\left(\cosh (\gamma )-\sinh (\gamma )\frac\partial{\partial\gamma}\right)e^{-\sqrt{\gamma  (\gamma +2 k)}}.
\end{split}
\end{equation}
Performing the inverse Laplace transform, we recover the result quoted in Ref.~\cite{Efetov1993},
\begin{equation}
P_\rho(s)=\sqrt{\frac{\gamma}{8\pi s^3} } e^{-\frac{\gamma}{2} \left(s+\frac{1}{s}\right)} \left(2 \cosh (\gamma )+\sinh (\gamma )
   \left(s+\frac{1}{s}-\frac{1}{\gamma }\right)\right).
\end{equation}
In this calculation, we have used that $Z[J_0+J_1]$ is independent of $t$, since for zero fermionic source, $J^{AR}_{BB}$ and  $J^{RA}_{BB}$ always appear multiplied by each other by symmetry (the same is true for the $FF$ component). This makes the integral over $t$ trivial. Interestingly, this result is also the kernel used to solve the non-linear sigma model in one dimension and on the Cayley tree \cite{Zirnbauer1986,Zirnbauer1986a,EfetovBook}.

From our splitting of the generating function into the standard saddle $Z_0$ and Andreev-Altshuler saddle $Z_1$ contributions (see Eq.~\eqref{eq_Z0_Z1}), we can analyze the contributions of each saddle $P_{\rho,0}$ and $P_{\rho,1}$ to the distribution, which read
\begin{equation}
\begin{split}
P_{\rho,0} (s)&= e^{\gamma}\sqrt{\frac{\gamma}{8\pi s^3} } e^{-\frac{\gamma}{2} \left(s+\frac{1}{s}\right)} \left(1+\frac12   \left(s+\frac{1}{s}-\frac{1}{\gamma }\right)\right),\\
P_{\rho,1} (s)&= e^{-\gamma}\sqrt{\frac{\gamma}{8\pi s^3} } e^{-\frac{\gamma}{2} \left(s+\frac{1}{s}\right)} \left(1-\frac12   \left(s+\frac{1}{s}-\frac{1}{\gamma }\right)\right).
\end{split}
\end{equation}
For large value of $\gamma$, the Andreev-Altshuler saddle contribution is exponentially suppressed. In the opposite limit, its contribution is however as important as that of the standard saddle (in particular, for $\gamma<\frac14$, $P_{\rho,0} (s)$ becomes negative for large enough $s$, and $P_{\rho,1} (s)$ is necessary to insure the positivity of the probability distribution).

\subsection{Probability distribution of a quantum dot conductance }
Another non-trivial result that can be obtained from the generating function is the probability distribution of the conductance of a quantum dot \cite{Prigodin1993}, given by ($\alpha$ is a positive number)
\begin{equation}
P_c(g)=\left\langle \delta\left(g+\alpha Q^{AR}_{BB}Q^{RA}_{BB}\right)\right\rangle_{J_0}.
\end{equation}
In this case, computing the Laplace transform is not sufficient, since $e^{k\alpha Q^{AR}_{BB}Q^{RA}_{BB}}$ is not a source term. The exponent can however be linearlized by using a complex variable $z$ as
\begin{equation}
\begin{split}
\tilde P_c(k)&=\int\frac{dz dz^*}{2i\pi}e^{-|z|^2}\left\langle e^{-\sqrt{k \alpha} \left(z Q^{AR}_{BB}+z^* Q^{RA}_{BB}\right)}\right\rangle_{J_0},\\
&=\int\frac{dz dz^*}{2i\pi}e^{-|z|^2}Z[J_0+J_2],
\end{split}
\end{equation}
with $\frac{dz dz^*}{2i\pi}=\frac{d{\rm Re}zd{\rm Im}z}{\pi}$ and 
\begin{equation}
(J_2)_{BB}=\sqrt{k \alpha}\begin{pmatrix}
0 & z \\
z^* & 0
\end{pmatrix},
\end{equation}
with the other  submatrices of $J_2$ being zero.\footnote{Here $(J_0+J_2)_{BB}$ is not of the form of Eq.~\eqref{eq_JBB}, but this does not compromise the convergence of the integrals.} This gives
\begin{equation}
\begin{split}
\tilde P_c(k)&=\int\frac{dz dz^*}{2i\pi}e^{-|z|^2-\sqrt{\gamma ^2+4 \alpha  k|z|^2}} \left( \cosh (\gamma )+\sinh (\gamma )
   \frac{\left(\gamma ^2+2 \alpha  k |z|^2\right)}{\gamma  \sqrt{\gamma ^2+4
   \alpha  k |z|^2}}\right),\\
   &=\left( \cosh (\gamma )-\sinh(\gamma )\left(\frac\alpha\gamma\frac\partial{\partial\alpha}+\frac\partial{\partial\gamma}\right)\right)\int\frac{dz dz^*}{2i\pi}e^{-|z|^2-\sqrt{\gamma ^2+4 \alpha  k|z|^2}}.
\end{split}
\end{equation}
Performing first the inverse Laplace transform, and then the integral over $z$, one obtains
\begin{equation}
\begin{split}
 P_c(g)&=\left( \cosh (\gamma )-\sinh(\gamma )\left(\frac\alpha\gamma\frac\partial{\partial\alpha}+\frac\partial{\partial\gamma}\right)\right) \frac{e^{-\gamma\sqrt{1+\frac{g}{\alpha}}}}{2\alpha}\left(\frac{\gamma}{1+\frac{g}{\alpha}}+\frac{1}{\left(1+\frac{g}{\alpha}\right)^{3/2}}\right),\\
 &=\frac{1}{2}\left( \cosh (\gamma )-\sinh(\gamma )\left(\frac{1-\lambda^2}{2\gamma\lambda}\frac\partial{\partial\lambda}+\frac\partial{\partial\gamma}\right)\right)\frac{\lambda^2-1}{g}e^{-\gamma\lambda}\left(\gamma\lambda^{-2}+\lambda^{-3}\right),
\end{split}
\end{equation}
with $\lambda=\sqrt{1+\frac g\alpha}$, which is equal to the result of \cite{Prigodin1993}, 
\begin{equation}
 P_c(g)=-\frac{1}{4\alpha\lambda}\frac\partial{\partial\lambda}\left(\frac{2}{\lambda}e^{-\gamma\lambda}\cosh (\gamma )+\left(1-\frac{1}{\gamma\lambda}+\frac{1}{\lambda^2}+\frac{1}{\gamma\lambda^3}\right)e^{-\gamma\lambda}\sinh (\gamma )\right).
 \end{equation}

Once again, one can separate the contribution of the two saddles, the standard saddle contribution coming from the $e^\gamma$ terms from the hyperbolic trigonometric functions and the Andreev-Altshuler saddle contribution coming from the $e^{-\gamma}$ terms. The behavior of the two terms is similar to those of the distribution of the density of state.

\section{Discussion \label{sec_concl}}

We have computed the generating function of a supersymmetric non-linear sigma model in the zero-dimensional (or zero-mode) limit, describing GUE random matrices, quantum chaotic systems, and small disordered metallic grains. Using the full range of invariance of the generating function, we were able to give its explicit form in terms of only three invariant functions of the source. Furthermore, we have been able to recover various non-trivial results found in the literature rather straightforwardly from this generating function.

One of the goals of our study was to better understand the functional dependence of the generating function, to aim for a functional renormalization group study of the Anderson transition. In the lattice formulation of the FRG, which is well suited to describe constrained field theories, see for instance \cite{Machado2010,Rancon2014b}, the zero-dimensional limit serves as an initial condition for the flow of the (scale-dependent) effective action $\Gamma$. Unfortunately, we have shown here that the effective action is not well defined for this supersymmetric model. Indeed, all physical values of the source ($J\propto \Lambda$) correspond to only one value of the average $\langle Q\rangle=\Lambda$, which does not allow for the construction of the effective action. This calls for more work to devise a FRG scheme that is compatible with the very specific functional structure of this supersymmetric model.

Another interesting direction would be the calculation of the generating function for other universality classes, such as the Gaussian Orthogonal Ensemble. In such cases, the semi-classical limit is not exact anymore, and it would be interesting to see how this would play a role. However, assuming that there is no other hidden invariance of the generating function, one should expect that the generating function is in principle a function of seven invariants (eight eigenvalues, minus one due to the shift-invariance), which might be challenging. A simpler calculation would be that of the generating function of the Circular Unitary Ensemble, following Ref.~\cite{HaakeBook}.

Finally, the calculation of the generating function in higher-dimensions would be of great interest. While the calculation for arbitrary space-dependent sources is completely out of reach, computing the generating function for constant sources might be doable at least in the one-dimensional case or on the Bethe lattice, by generalizing the methods of Refs. \cite{Zirnbauer1986a,Khalaf2017,Khalaf2017a}.

\subsection*{ACKNOWLEDGMENTS}
A.R. warmly thanks C. Tian for numerous interesting discussions and suggestions. We thank F. Franchini for discussions at the early stage of this work, and M. Kazalicki for pointing out the relevance of one-parameter Lie groups.  A.R. thanks the Institute of Physics of Zagreb for its hospitality, where part of this work was done.  I.B. acknowledges the support of the Croatian Science Foundation Project No. IP-2016-6-7258, IP-2016-6-3347 and the QuantiXLie Centre of Excellence, a project cofinanced by the Croatian Government and European Union through the European Regional Development Fund - the Competitiveness and Cohesion Operational Programme (Grant KK.01.1.1.01.0004). This study was supported by the French government through the Programme Investissement d’Avenir (I-SITE ULNE/ANR-16-IDEX-0004 ULNE) managed by the Agence Nationale de la Recherche.

\appendix

\section{Computation of $\Q[J]$ \label{app_Q}}

The average $\Q[J]=\langle Q\rangle_J$ is obtained from the generating function $Z[J]$ using
\begin{equation}
\Q_{AB}[J]=\frac{1}{Z[J]}\frac{\delta Z}{\delta \tilde J_{AB}},
\end{equation}
where $A,B$ are collective (AR and FB) indices, and $\tilde J_{AB}=(-1)^{|B|+1}J_{BA}$, with $|B|=1$ if it corresponds to a fermionic index and $0$ else. 

Using that $\frac{\delta j_n}{\delta \tilde J_{AB}}=n (J^{n-1})_{AB}$, and that $Z[J]$ depends on $j_n$, on observe that $\Q[J]$ is diagonal when $J$ is. One can therefore focus on the diagonal case to compute the $\Q$ invariants. Thanks to the linear constraint $\Str(Q)=0$ (or equivalently to the shift invariance of $Z[J]$) one of course explicitly find that $\Str(\Q[J])=0$. For diagonal sources $L$, it is therefore convenient to parametrize the (diagonal) average as $\Q[L]=M$, with
\begin{equation}
M[L]= M_F e_F+ M_B e_B+ M_{\mathbb{1}}\mathbb{1},
\end{equation}
with $e_F$ and $e_B$ given in Eq.~\eqref{eq_basis}. An explicit calculation gives

\begin{equation}
\begin{split}
M_F & =\frac{2 \left(I_F 
   \left(I_B^2+I_F^2-i_1\right)\cosh (I_F)+ \left(i_1-I_B^2+(2 I_B+1)
   I_F^2\right)\sinh (I_F)\right)}{I_F \left(2 I_B I_F \cosh
   (I_F)+   \left(I_F^2+I_B^2-i_1\right)\sinh (I_F)\right)},\\
M_B &= \frac{4
   I_B^2 I_F \cosh (I_F)+2  \left(-i_1 (I_B+1)+(I_B-1) I_B^2+(I_B+1) I_F^2i_1\right)\sinh (I_F)}{I_B \left(2 I_B I_F \cosh
   (I_F)+   \left(I_F^2+I_B^2-i_1\right)\sinh (I_F)\right)},\\
M_{\mathbb{1}} &=-\frac{8 \sqrt{i_1} \sinh (I_F)}{2 I_B I_F \cosh
   (I_F)+
   \left(I_F^2+I_B^2-i_1\right)\sinh (I_F)}.
\end{split}
\end{equation}
From this, $\Str(\Q[J]^n)=\Str(M[L]^n)$ is readily obtained for arbitrary source. 
Note that if $j_1=\pm (I_F-I_B)$, one finds $M_F=M_B$, which translates into $\Str(\Q[J]^n)=0$.


%

\end{document}